\documentclass[journal]{IEEEtran}

\usepackage[numbers,sort&compress]{natbib}
\usepackage{amsmath}
\usepackage{amssymb}
\usepackage{graphicx}
\usepackage{epstopdf}

\graphicspath{{fig/}}
\DeclareGraphicsExtensions{.eps}
\usepackage{epstopdf}

\begin{document}
%
\title{Reduce the Complexity of List Decoding of Polar Codes by Tree-Pruning}
%
%
%
\author{Kai~Chen,
        Bin~Li,
        Hui~Shen,
        Jie~Jin,
        and David~Tse, \IEEEmembership{Fellow,~IEEE}
\thanks{This work was supported by the Science and Technology Project of Shenzhen (No. JSGG20141020145124600).}
\thanks{K. Chen, B. Li, J. Jin, and H. Shen are with the Communications Technology Research Lab., Huawei Technologies, Shenzhen, P. R. China (e-mail: kaichen@ieee.org).}
\thanks{D. Tse is with the Dept. of Electrical Engineering, Stanford University, CA 94305-9515, USA (e-mail: dntse@stanford.edu).}
}

%

\maketitle

\newtheorem{theorem}{Theorem}
\newtheorem{example}{Example}
\newtheorem{algorithm}{Algorithm}
\begin{abstract}
Polar codes under cyclic redundancy check aided successive cancellation list (CA-SCL) decoding can outperform the turbo codes and the LDPC codes when code lengths are configured to be several kilobits.
In order to reduce the decoding complexity, a novel tree-pruning scheme for the \mbox{SCL/CA-SCL} decoding algorithms is proposed in this paper.
In each step of the decoding procedure, the candidate paths with metrics less than a threshold are dropped directly to avoid the unnecessary computations for the path searching on the descendant branches of them.
Given a candidate path, an upper bound of the path metric of its descendants is proposed to determined whether the pruning of this candidate path would affect frame error rate (FER) performance.
By utilizing this upper bounding technique and introducing a dynamic threshold, the proposed scheme deletes the redundant candidate paths as many as possible while keeping the performance deterioration in a tolerant region, thus it is much more efficient than the existing pruning scheme.
With only a negligible loss of FER performance, the computational complexity of the proposed pruned decoding scheme is only about $40\%$ of the standard algorithm in the low signal-to-noise ratio (SNR) region (where the FER under CA-SCL decoding is about $0.1 \sim 0.001$), and it can be very close to that of the successive cancellation (SC) decoder in the moderate and high SNR regions.

\end{abstract}


\section{Introduction}

\IEEEPARstart{P}{olar} codes have been proven to achieve the symmetric capacity on binary-input discrete memoryless channels under a low-complexity successive cancellation (SC) decoding algorithm \cite{pcode}.
Although the polar codes asymptotically achieve the channel capacity, the performance under SC decoding is unsatisfying when the code length is of the order of kilobits.
Several alternative decoding schemes have been proposed to improve the finite-length performance of polar codes, such as successive cancellation list (SCL) \cite{scl}, successive cancellation stack (SCS)\cite{scs} and belief propagation (BP) \cite{bp} decoding algorithms.
It is reported that polar codes under the CRC-aided SCL/SCS (CA-SCL/SCS) decoding algorithms can achieve a better frame error rate (FER) performance than the LDPC and turbo codes when the code lengths are configured to several kilobits \cite{cascl}\cite{acascl}\cite{cascs}.
Therefore, polar coding is believed to be a competitive candidate in future communication systems.

Since the CA-SCS decoding requires a large stack to store the candidate paths which leads to a high space complexity, CA-SCL decoding algorithm is of more interest \cite{list1}\cite{list2}\cite{list3}.
Nevertheless, to achieve competitive performance against LDPC or turbo codes, a moderate-sized list is required in CA-SCL decoding.
In that case, the computational complexity of the CA-SCL decoder is still high.

As stated in \cite{sch}, SCL decoding can be regarded as a path searching procedure on the code tree.
In order to reduce the complexity of SCL decoding, tree-pruning technique is exploited by avoiding unnecessary path searching operations \cite{vtc}.
In order to keep the FER loss in an acceptable region, \cite{vtc} computes the pruning threshold in a very conservative way.
Only the candidate paths with metrics much less than the maximum one are pruned.
It works well when the signal-to-noise ratio (SNR) is high, where the metric of the correct path is usually much larger than the others.
However, this existing pruning technique is no longer efficient when working in the relative low SNR region where the FER under \mbox{CA-SCL} decoding is about $0.1 \sim 0.001$, while it is exactly the operating regime for cellular networks.

In this paper, we propose to compute the threshold using the sum of the survival path metrics.
To evaluate how much a pruned candidate path would affect FER performance, we propose a metric upper bound of its descendants.
Utilizing this upper bounding technique, a dynamic threshold is further proposed.
The proposed scheme deletes the redundant candidate paths as many as possible while keeping the performance deterioration in a tolerant region, thus it is much more efficient than the existing pruning scheme.

The remainder of the paper is organized as follows.
\mbox{Section \ref{sec_pre}} reviews the basics of polar coding.
\mbox{Section \ref{sec_sch}} describes the proposed tree-pruning scheme for SCL decoding. A path metric upper bound of the descendants of some given candidate path and a dynamic threshold configuration method are proposed.
\mbox{Section \ref{sec_sim}} provides the performance and complexity analysis based on the simulation results.
Finally, \mbox{Section \ref{sec_con}} concludes the work.

\section{Preliminaries}
\label{sec_pre}

\subsection{Notation Convention}

In this paper, we use calligraphic characters, such as $\mathcal{X}$ and $\mathcal{Y}$, to denote sets, and $|\mathcal{X}|$ to denote the number of elements in $\mathcal{X}$.
We write the Cartesian product of $\mathcal{X}$ and $\mathcal{Y}$ as $\mathcal{X} \times \mathcal{Y}$, and write the $n$-th Cartesian power of $\mathcal{X}$ as ${\mathcal{X}}^n$.
Further, we write $\mathcal{Y} \backslash \mathcal{X}$ to denote the subset of $\mathcal{Y}$ with elements in $\mathcal{X}$ excluded.

We use notation $v_1^N$ to denote a $N$-dimension vector $\left(v_1, v_2, \cdots, v_N \right)$ and $v_i^j$ to denote a subvector $\left(v_i, v_{i+1},\cdots, v_{j-1}, v_j\right)$ of $v_1^N$, $1\leq i,j \leq N$.
Particularly when $i>j$, $v_i^j$ is a vector with no elements in it and the empty vector is denoted by $\phi$.
We write $v_{1,o}^N$ to denote the subvector of $v_1^N$ with odd indices ($a_k: 1 \leq k \leq N$; $k$ is odd).
Similarly, we write $v_{1,e}^N$ to denote the subvector of $v_1^N$ with even indices ($a_k: 1 \leq k \leq N$; $k$ is even).
For example, for $v_1^4$, $v_{2}^3=(v_2,v_3)$, $v_{1,o}^4=(v_1,v_3)$ and $v_{1,e}^4=(v_2,v_4)$.
Further, given a index set $\mathcal{A}$, $v_{\mathcal{A}}$ denote the subvector of $v_1^N$ which consists of $v_i$s with $i \in \mathcal{A}$.

\subsection{Polar Coding and SC Decoding}

We are given a binary-input memoryless channel $W: \mathcal{X} \to \mathcal{Y}$ with input alphabet $\mathcal{X}=\left\{0,1\right\}$ and output alphabet $\mathcal{Y}$, the channel transition probabilities are $W\left(y|x\right)$, $x\in \mathcal{X}$, $y\in \mathcal{Y}$.

For code length $N=2^n$, $n = 1,2,\cdots$, and information length $K$, i.e. code rate $R=K/N$, polar coding over $W$ proposed by Ar{\i}kan can be described as follows:

After channel combining and splitting operations on $N$ independent uses of $W$, we get $N$ successive uses of synthesized binary input channels $W_N^{(i)}$, $i=1,2,\cdots,N$, with transition probabilities
\begin{equation}
\label{equ_polarized_channels}
     {W}_N^{(i)}(y_1^N, u_1^{i-1}|u_i)=\sum\limits_{u_{i+1}^{N} \in \mathcal{X}^{N-i}}{\frac{1}{2^{N-1}}{W}_N(y_1^N|u_1^N)}
\end{equation}
\noindent{where}
\begin{equation}
\label{equ_polarized_channels2}
    {W}_N(y_1^N|u_1^N)=\prod\limits_{{i}=1}^{N}{W(y_{i}|x_{i})}
\end{equation}
and the source block $u_1^N$ are supposed to be uniformly distributed in ${\left\{0,1\right\}}^{N}$.

The reliabilities of the polarized channels $\left\{W_N^{(i)}\right\}$ can be evaluated by using density evolution \cite{de}, and is usually more evaluated efficiently by calculating Bhattacharyya parameters \cite{pcode} for binary erasure channels (BECs) or by using Gaussian approximation \cite{ga} for binary-input AWGN (BIAWGN) channels.

To transmit a message block of $K$ bits, the $K$ most reliable polarized channels $\left\{W_N^{(i)}\right\}$ with indices $i \in \mathcal{A}$ are picked out for carrying these information bits; a fixed bit sequence called frozen bits are transmitted over the others.
The index set $\mathcal{A} \subseteq \left\{1, 2, \cdots, N\right\}$ is called the information set and $|\mathcal{A}|=K$, and its complement set which is denoted by $\mathcal{A}^c$ is called the frozen set.

As mentioned in \cite{pcode}, polar codes can be decoded using successive cancellation (SC) decoding algorithm.
In \cite{sch}, it is further described as a path searching procedure on a decoding tree.
The metric of a decoding path $u_1^i$ can be measured using \emph{a posteriori} probability
\begin{equation}
\label{equ_app}
P_N^{(i)}{\left(\hat{u}_1^i|y_1^N\right)}=
\begin{cases}
0 & \text{if } i \in \mathcal{A}^c \text{ and } \hat{u}_i \neq u_i \\
\frac{{W}_N^{(i)}(y_1^N, u_1^{i-1}|u_i)}{2P\left(y_1^N\right)} & \text{otherwise} \\
\end{cases}
\end{equation}
When $\hat{u}_i$ is not a wrong frozen bit, the above path metric can be recursively computed as
\begin{equation}
\label{equ_app_recursive1}
    \begin{aligned}
    & P_{2N}^{\left( 2i-1 \right)}\left( \hat{u}_{1}^{2i-1}\left| y_{1}^{2N} \right. \right) \qquad \qquad \qquad \qquad \qquad\\
    & =\sum\limits_{{\hat{u}_{2i}\in\{0,1\}}}{P_{N}^{\left( i \right)}\left( \hat{u}_{1,o}^{2i}\oplus \hat{u}_{1,e}^{2i}\left| y_{1}^{N} \right. \right)
    \cdot P_{N}^{\left( i \right)}\left( \hat{u}_{1,e}^{2i}\left| y_{N+1}^{2N} \right. \right)}
    \end{aligned}
\end{equation}
\begin{equation}
\label{equ_app_recursive2}
    \begin{aligned}
    & P_{2N}^{\left( 2i \right)}\left( \hat{u}_{1}^{2i}\left| y_{1}^{2N} \right. \right) \qquad \qquad \qquad \qquad \qquad \qquad\\
    & =P_{N}^{\left( i \right)}\left( \hat{u}_{1,o}^{2i}\oplus \hat{u}_{1,e}^{2i}\left| y_{1}^{N} \right. \right)
    \cdot P_{N}^{\left( i \right)}\left( \hat{u}_{1,e}^{2i}\left| y_{N+1}^{2N} \right. \right)\qquad \quad
    \end{aligned}
\end{equation}
where $n\ge 0$, $N={{2}^{n}}$, $1\le i\le N$.

Thus, SC decoding can be described as a greedy search algorithm on the code tree.
In each level, only the one of two descendants with larger path metric is selected for further expansion.

\begin{equation}
\label{equ_sc}
\hat{u}_i=
\begin{cases}
 {h_i}\left(y_1^N, \hat{u}_1^{i-1}\right) & i \in \mathcal{A}\\
 u_i & i \in \mathcal{A}^c\\
\end{cases}
\end{equation}
where
\begin{equation}
\label{equ_sc_h}
{h_i}\left(y_1^N, \hat{u}_1^{i-1}\right) =
\begin{cases}
 0 & \text{if }{P_N^{(i)}\left(\hat{u}_1^{i} | y_1^N\right)} \ge {P_N^{(i)}\left(\hat{u}_1^{i} | y_1^N \right)}\\
 1 & \text{otherwise}\\
\end{cases}
\end{equation}

\subsection{Improved SC Decoding Algorithms}

The performance of SC is limited by the bit-by-bit decoding strategy. Whenever a bit is wrongly determined, there is no chance to correct it in the rest of the decoding procedure.

Theoretically, the performance of the maximum \emph{a posteriori} probability (MAP) decoding (or equivalently ML decoding, since the inputs are assumed to be uniformly distributed) can be achieved by traversing all the $N$-length decoding paths in the code tree.
But this brute-force search takes exponential complexity and is impossible to be implemented for practical code lengths.

Two improved decoding algorithms, SCL decoding and SCS decoding, are proposed in \cite{scl} and \cite{scs}.
Both of these two algorithms allow more than one edge to be explored in each level of the code tree.
During the SCL(SCS) decoding, a set of candidate paths are obtained and stored in a list(stack).
Combining the ideas of SCL and SCS, a decoding algorithm named successive cancellation hybrid (SCH) is proposed in \cite{sch}, which can achieve a better trade-off between computational complexity and space complexity.
Moreover, with the help of CRC codes, polar codes decoded by these improved SC decoding algorithms are found to be capable of achieving the same or even better performance than turbo codes or LDPC codes \cite{cascl} \cite{cascs} \cite{acascl}.

\begin{figure}
\centering{
\includegraphics[width=0.8\columnwidth]{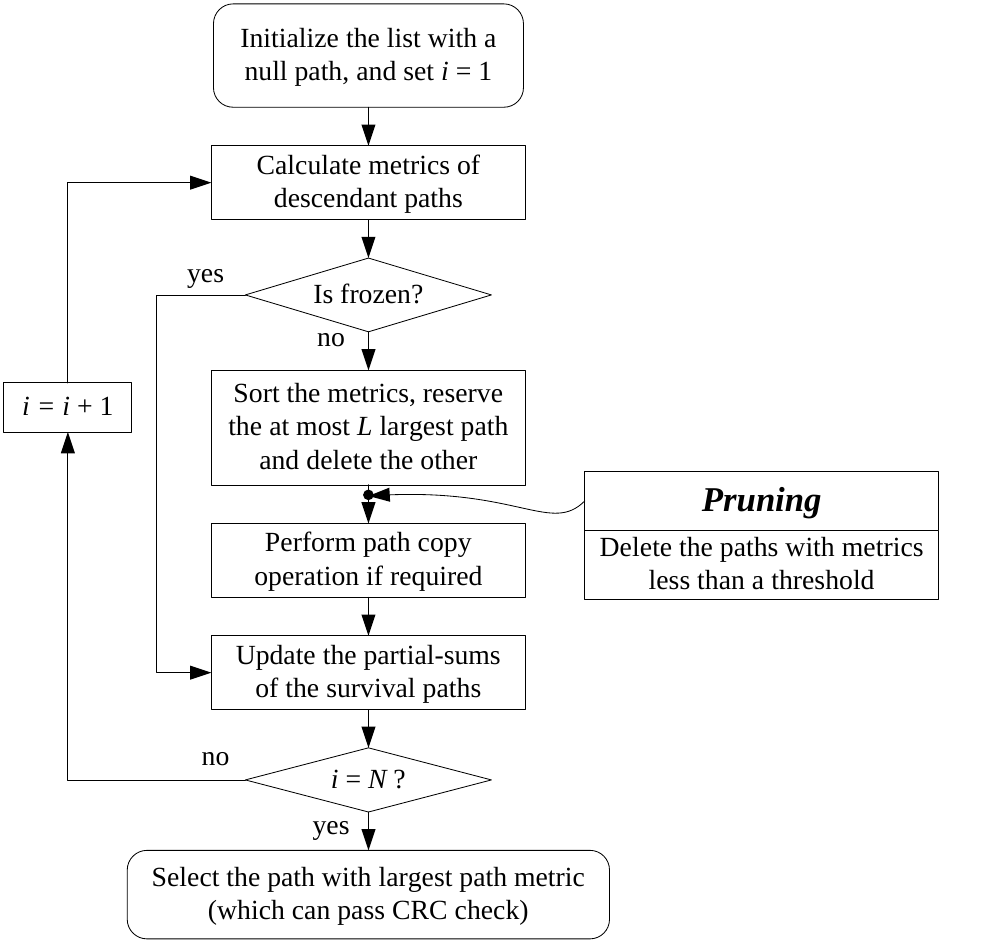}}
\caption{The flow chart of the pruned SCL/CA-SCL decoding.}
\label{fig_scl}
\end{figure}

Among these existing improved SC decoding algorithms, benefitting from the limited requirement for the memory, \mbox{(CA-)SCL} decoding is the most interesting for hardware implementation \cite{list1} \cite{list2} \cite{hw_list1} \cite{hw_list2}.
As shown in Fig. \ref{fig_scl}, the processing loop of the standard SCL/CA-SCL decoding is as follows:

\begin{enumerate}

\item[S1)]  For each candidate path, calculate the path metrics of its descendant paths;

\item[S2)]  Sort the metrics, and reserve at most $L$ paths with the larger metrics and delete the others;

\item[S3)]  If any two of the survival paths share the same parent node, then a copy operation is performed to create separate working spaces for these two paths;

\item[S4)]  For each survival path, update the \mbox{partial-sum} recursively;

\item[S5)]  The above loop is processed until the length of candidate paths reach $N$. The candidate path with the largest path metric (when CRC embedded, the candidates which cannot pass CRC are dropped directly) is picked out for the final decision.

\end{enumerate}
\rightline{$\blacksquare$}

%

\section{Tree-Pruning Scheme for SCL/CA-SCL Decoding Algorithm}
\label{sec_sch}

\subsection{The Proposed Pruning Scheme}

In order to reduce the computational complexity of SCL decoding, a pruning operation is added after the sorting operation (as shown in Fig.\ref{fig_scl}).
If the metric of some candidate path is less than a threshold, it will be directly deleted to avoid redundant path expansions and copy operations.

In this paper, we propose to use the path metric sum of the (maximum) $L$ survival candidate paths after sorting:
while decoding the $i$-th bit, the metrics of the survival paths is $\left\{P^{(i)}_j\right\}$, where $j \in \mathcal{L}_i$ is index set of the survival paths in the list after sorting operation, $1 \le \left|\mathcal{L}_i\right| \le L$;
If the following inequality holds for some $j \in \mathcal{L}_i$, the corresponding path is then deleted,
\begin{equation}
\label{eqn_prune}
P^{(i)}_j < \alpha_i \cdot \sum_{k=1}^{\left| \mathcal{L}_i \right|}{P^{(i)}_k}
\end{equation}
where $0 \le \alpha_i < 1$. Particularly, if $\alpha_i=0$, then no pruning is performed when decoding this $i$-th bit.
In the following part of this section, we'll discuss how to choose the value of $\{\alpha_i\}$.

\subsubsection{Performance Deterioration}

Suppose that the correct path is still in the list after the sorting operation during decoding the $i$-th bit.
The probability of that the $j$-th candidate is the correct path (i.e., the performance loss of deleting this path) is computed as
\begin{equation}
\label{eqn_pde}
\Pr\left\{j\text{-th path is the correct one}\right\} = \frac{P_j^{(i)}}{\sum_{k=1}^{\left| \mathcal{L}_i \right|}{P_k^{(i)}}}
\end{equation}

\subsubsection{Statistical Threshold Configuration}

Given a specific polar code, the channel property, and a tolerant FER performance loss $P_\text{tol}$, the most direct way to configure $\alpha_i$ is through Mote Carlo simulation.

Initially, set $\alpha_i=1$ and simulate using standard (CA-)SCL decoding.
During decoding the $i$-th bit in each frame, the ratio of the metric of the correct path (until the $i$-th bit) $P_c^{(i)}$ and the sum metric of the survival paths in the list is recorded; If the final decoding result is correct and the ratio is less than $\alpha_i$, then update $\alpha_i$ with this ratio, i.e.,
\begin{equation}
\alpha_i = \min \left( \alpha_i, \frac{P_c^{(i)}}{\sum_{k=1}^{L}{P_k^{(i)}}} \right)
\end{equation}
When the amount of simulated frame is large enough, the pruning operation based on (\ref{eqn_prune}), the FER performance loss can be very small.

\subsection{Dynamic Threshold Configuration}

The Monte Carlo configuration is dependant on the specific SNR, code length, and code rate. Thus, it's quite difficult to use for practical application.
For polar codes, the reliability of the polarized channels can be evaluated using Gaussian approximation \cite{ga} or some other techniques; in other words, the probability density functions (PDFs) of the LLRs which corresponding to the receiving bits (conditioned on that the previous bits are correctly decoded) can be a priori information to the decoder.
In this subsection, we present a method to estimate the performance loss brought by pruning using these LLR distributions; and then, a dynamic threshold configuration method is proposed.
Using the proposed thresholding method, the pruned (CA-)SCL decoding can fully utilize the tolerant performance deterioration and thus lower the computational complexity.

\subsubsection{Path Metric Upper Bounds}

\begin{figure}
\centering{
\includegraphics[width=0.85\columnwidth]{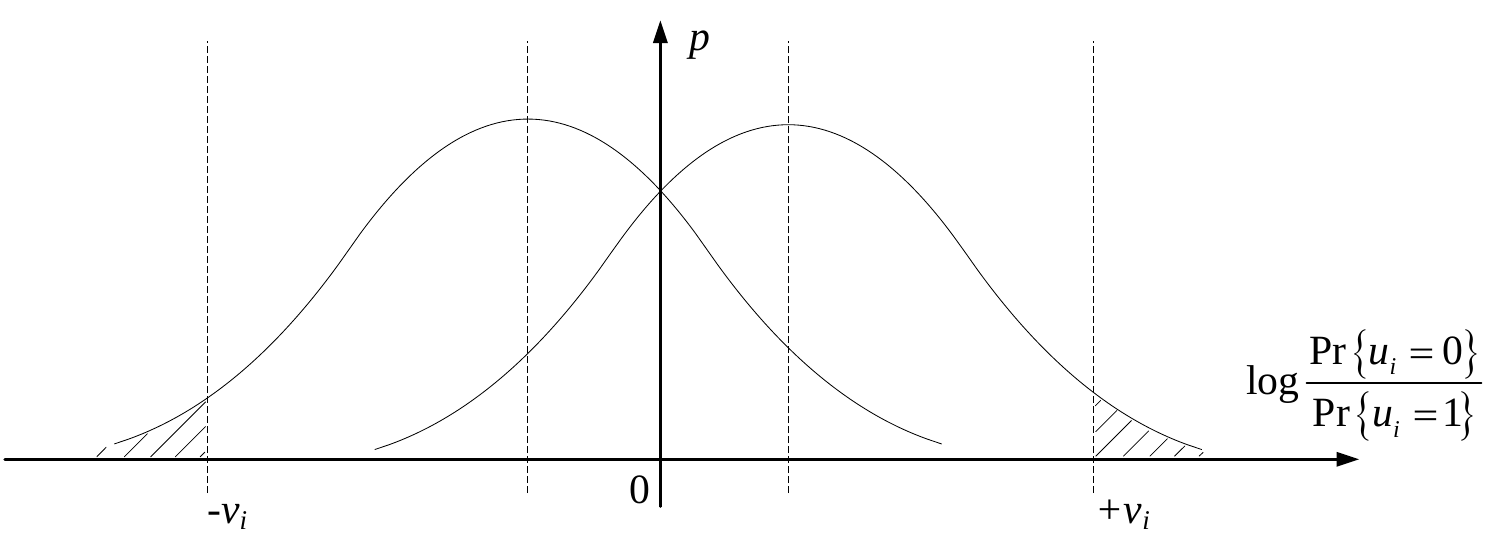}}
\caption{The probability density function of LLR.}
\label{fig_llrpdf}
\end{figure}

\begin{figure}
\centering{
\includegraphics[width=0.85\columnwidth]{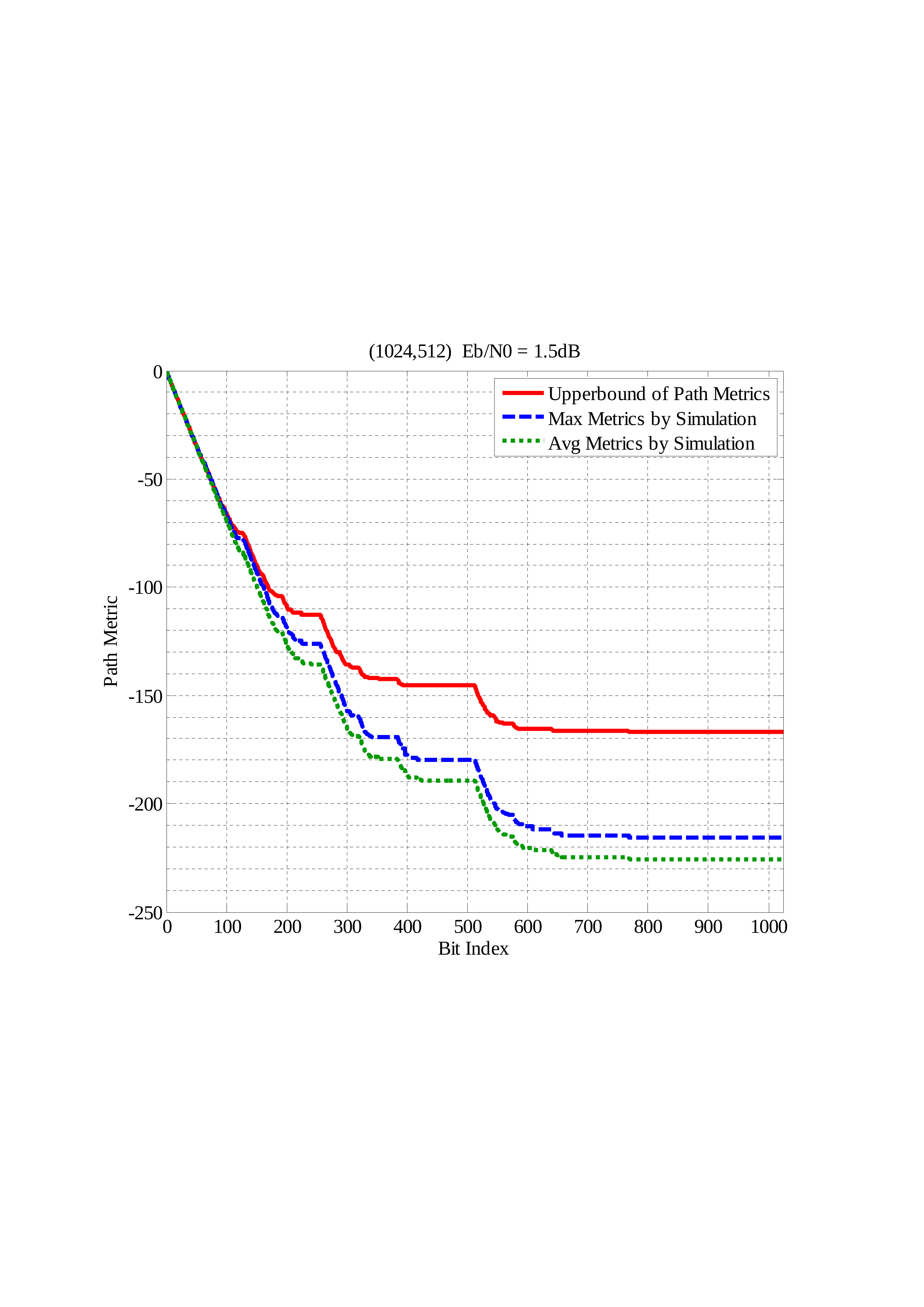}}
\caption{The upper bound of path metric.}
\label{fig_metricupperbound}
\end{figure}

The LLR PDFs can be obtained by using density evolution or Gaussian approximation \cite{ga}.
Based on the PDF corresponding to a bit $u_i$, we can define a LLR region, such that the probability of the corresponding LLR takes values in $\left[-l_i, l_i\right]$ is larger than a pre-defined small probability $P_\text{llr}$,
\begin{equation}
\Pr\left\{-l_i \le \log \frac{\Pr\{u_i=0\}}{\Pr \{u_i=1\}} \le l_i\right\} \ge 1-P_\text{llr}
\end{equation}
Therefore, when decoding the $i$-th bit, if one candidate path has metric $P^{(i)}$, the metric of any its descendant path $P^{(j)}$ at the $j$-th level has an upper bound,
\begin{equation}
\label{eqn_metricmax}
P^{(j)}_\text{ub} = P^{(i)} \cdot \prod_{k=i+1}^{j}(\frac{e^{l_i}}{1+e^{l_i}}) \ge P^{(j)}
\end{equation}
where $1 \le i < j \le N$.

Note that for bit index $k \in [i, j]$, every bit effects the value of $P_\text{ub}^{(j)}$ to some extent, no matter it's an information bit or a frozen bit.
Specifically, for an information bit with relatively high reliability, i.e., with a large $l_i$, its impact on $P_\text{ub}^{(j)}$ is considered negligible;
for a frozen bit, since the value of $l_i$ is relatively smaller, its impact on $P_\text{ub}^{(j)}$ is more significant.

Fig. \ref{fig_metricupperbound} gives the simulation result of a $(1024, 512)$ polar code under BIAWGNC with SNR $1.5$dB.
The decoding algorithm is \mbox{CA-SCL} with $L=32$.
The maximum and average values of the path metric during decoding each bit are recoded.
To guarantee the inequality (\ref{eqn_metricmax}) holds with probability larger than $1-10^{-9}$, we set $P_\text{llr}=\frac{10^{-9}}{N}$.
As shown in the figure, the simulation data is well bounded by (\ref{eqn_metricmax}).

\subsubsection{Threshold Computation}

In this subsection, we propose a new threshold computation method which can fully utilize the \mbox{pre-defined} tolerant FER performance loss $P_\text{tol}$.

\begin{figure}
\centering{
\includegraphics[width=0.95\columnwidth]{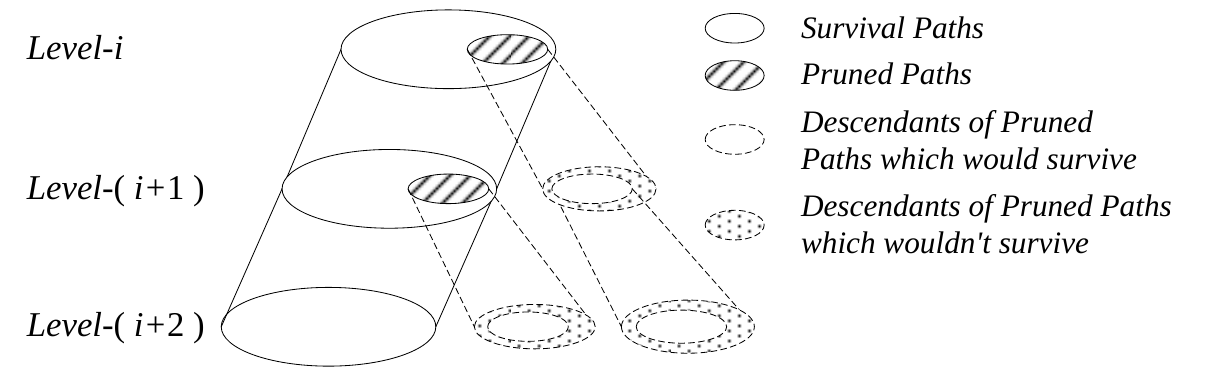}}
\caption{A graphic demonstration of the pruned paths which would survive.}
\label{fig_ploss}
\end{figure}

As previously stated, pruning operation during decoding $u_i$ will cause some FER performance loss;
When expansion at level-$(i+1)$ on the code tree, the loss brought by the pruned path at level-$i$ is accumulated, i.e., the paths which cause performance loss during decoding $u_{i+1}$ include not only the newly pruned paths but also the descendants of the pruned paths at level-$i$.
Thus, when decoding at level-$i$ on the code tree, the FER performance loss is computed based on both the newly pruned paths at level-$i$ and the descendants of all the previously pruned paths which would be in the list.
A graphic illustration is given in Fig. \ref{fig_ploss}.

To estimate the FER loss brought by the pruning operations, the pruned path should be recorded.
Let $\mathcal{S}_i$ be the \emph{active} pruned path during decoding the first $i$ bits, i.e., $u_1^i$.
For each pruned path $k \in \mathcal{S}_i$, the level index $t_k$ when it is pruned, along with the corresponding path metric $p_k$ and the estimated performance loss $q_k$ which is computed using (\ref{eqn_pde}), is recorded.
Obviously, $t_k \le i$.
Based on $\left(t_k, p_k, q_k\right)$, the maximum metric value at the $i$-th level of the descendants of the pruned path $k$ can be computed using $P_\text{ub}^{(i)}$ in (\ref{eqn_metricmax}),
\begin{equation}
Z_{k}^{(i)} = p_k - P^{(t_k)}_\text{ub} + P^{(i)}_\text{ub}
\end{equation}

The performance loss $P_\text{de}^{(i)}$ which is brought by the pruning operations during decoding the first $i$ bits is evaluated as follows:

\begin{enumerate}

\item[S1)]  Find the survival paths in the list $\mathcal{L}_i$ which are with metrics larger than the maximum $Z_k^{(i)}$,
\begin{equation}
\mathcal{L}'_i = \left\{ j\left| j \in \mathcal{L}_i,P_j^{(i)} \ge \max_{k \in \mathcal{S}_i}{Z_k^{(i)}} \right. \right\} \subseteq \mathcal{L}'_i
\end{equation}
the number of these found paths is $\left| \mathcal{L}'_i \right|$;

\item[S2)]   Find $(L-\left| \mathcal{L}'_i \right|)$ pruned records with indices $\mathcal{S}'_i \subseteq \mathcal{S}_i$ which has the larger estimated performance losses, i.e., for any $k \in \mathcal{S}'_{i}$ and $k' \in \mathcal{S}_{i} \backslash \mathcal{S}'_{i}$, we have $q_k \ge q_{k'}$, where $|\mathcal{S}'_i|=L-\left| \mathcal{L}'_i \right|$.

\item[S3)]  The performance loss $P_\text{de}^{(i)}$ is upper bounded by
\begin{equation}
P_\text{de}^{(i)} \le \sum _{k \in \mathcal{S}'_i} {q_k}
\end{equation}

\end{enumerate}
\rightline{$\blacksquare$}

The threshold $\alpha_i$ is determined by the tolerant performance loss $P_\text{tol}$ and the loss introduced in the previous decoding process $P_\text{de}^{(i-1)}$,
\begin{equation}
\alpha_i = \frac{\sum _{j \in \mathcal{L}_i \backslash \mathcal{R}_i} {P_j^{(i)}}} {\sum _{j \in \mathcal{L}_i} {P_j^{(i)}}}
\end{equation}
where index set $\mathcal{R}_i$ indicates the candidates to be pruned and is the largest subset of $\mathcal{L}_i$ which satisfies
\begin{equation}
\sum _{j \in \mathcal{R}_i} P_j^{(i)} \le \left(P_\text{tol}-P_\text{de}^{(i-1)}\right) \cdot \sum _{j \in \mathcal{L}_i} P_j^{(i)}
\end{equation}

After the pruning, the set of pruned records $\mathcal{S}_{i}$ is updated as follows:

\begin{enumerate}

\item[S1)] Combing $\mathcal{S}_{i-1}$ and the newly pruned paths which are induced by $\mathcal{R}_i$, the obtained temporary index set is denoted by $\mathcal{T}_i$;

\item[S2)] Find the $L$ pruned records with largest losses in $\mathcal{T}_i$, the result indices form the set $\mathcal{T}'_i$, i.e., $\mathcal{T}'_i \subseteq \mathcal{T}_i$, for any $k \in \mathcal{T}'_i$ and $k' \in \mathcal{T}_i \backslash \mathcal{T}_i$, we have $q_k \ge q_{k'}$.

\item[S3)] The minimum value of the metric upper bounds of the pruned records in $\mathcal{T}'_i$ is
\begin{equation}
Z_\text{min} = \min _{k \in \mathcal{T}'_i} {Z_k^{(i)}}
\end{equation}

\item[S4)] $\mathcal{S}_{i}$ is obtained by inactivating all the records in $\mathcal{T}_i$ with estimated metric less than $Z_\text{min}$
\begin{equation}
\mathcal{S}_i = \left\{ k \left| k \in \mathcal{T}_i, Z_k^{(i)} \ge Z_\text{min} \right. \right\}
\end{equation}

\end{enumerate}
\rightline{$\blacksquare$}
Note that, initially, $\mathcal{S}_{0}=\emptyset$.

\subsection{Complexity}

The complexity of (CA-)SCL decoding consists of three parts:
the path extension (includes the updating of path metrics (\ref{equ_app_recursive1}) (\ref{equ_app_recursive2}) and the partial-sums), path metric sorting, path copy, and \mbox{partial-sum} updating.

Applying pruning, many redundant path extensions along with path copies are avoided.
Since the computational complexity to obtain a length-$N$ path is $O(N \log N)$ \cite{sch}, and in the best case only one path is preserved in the list, thus the computational complexity is reduced by $O(LN\log N)$.
However, calculating the threshold itself introduces additional compactions. For each information bit, the metrics of the survival paths and the pruned paths are added up to compute the threshold, thus the complexity increases with $O(LN)$.
Thus, the computational complexity can be reduced by order of $O(LN\log N)$ if the $P_\text{tol}$ is set to a proper value.

Moreover, when one of the two descendants of a single parent path is pruned, there is no longer need for the path copy operation.
In fact, it is the usual case especially when the corresponding polarized channel is with high reliability.
Therefore, the number of required path copies is also reduced.

As to the path metric sorting, the least reliable paths are required to be picked out when computing the threshold, so the pruning does not reduce the sorting complexity.

\section{Simulation Results}
\label{sec_sim}

In this section, we analyze the performance of the proposed pruned (CA-SCL) decoding algorithm via simulation.
The simulated polar code has code length $N=1024$ and the code rate $R=1/2$, which is constructed under $E_b/N_0=1.5\text{dB}$ using Gaussian Approximation \cite{ga}.
The information block is assumed to have $16$ embedded CRC bits, and CA-SCL decoding is applied.

\begin{figure}
\centering{
\includegraphics[width=0.85\columnwidth]{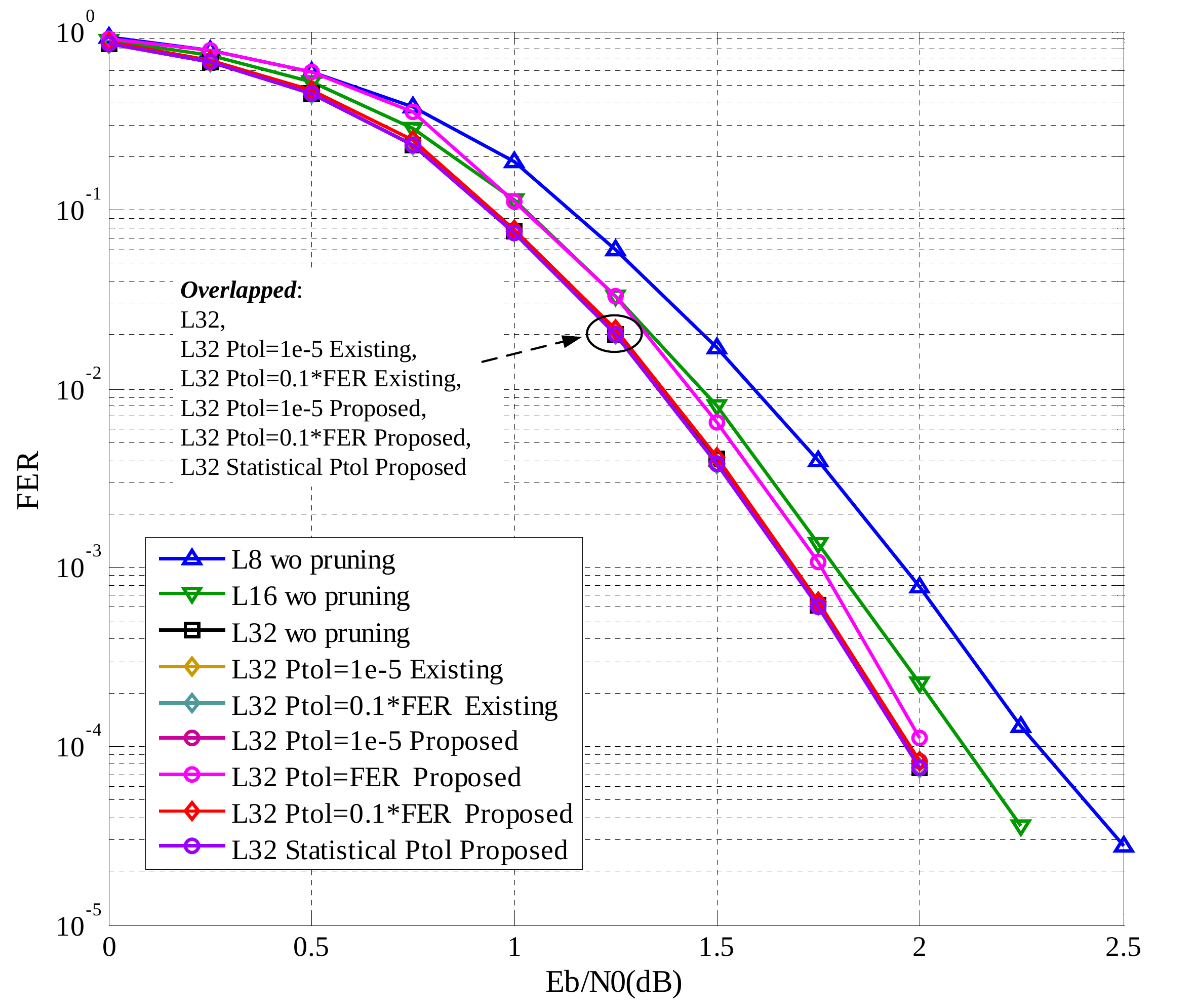}}
\caption{FER performances under different decoding schemes.}
\label{fig_bler}
\end{figure}

\begin{figure}
\centering{
\includegraphics[width=0.85\columnwidth]{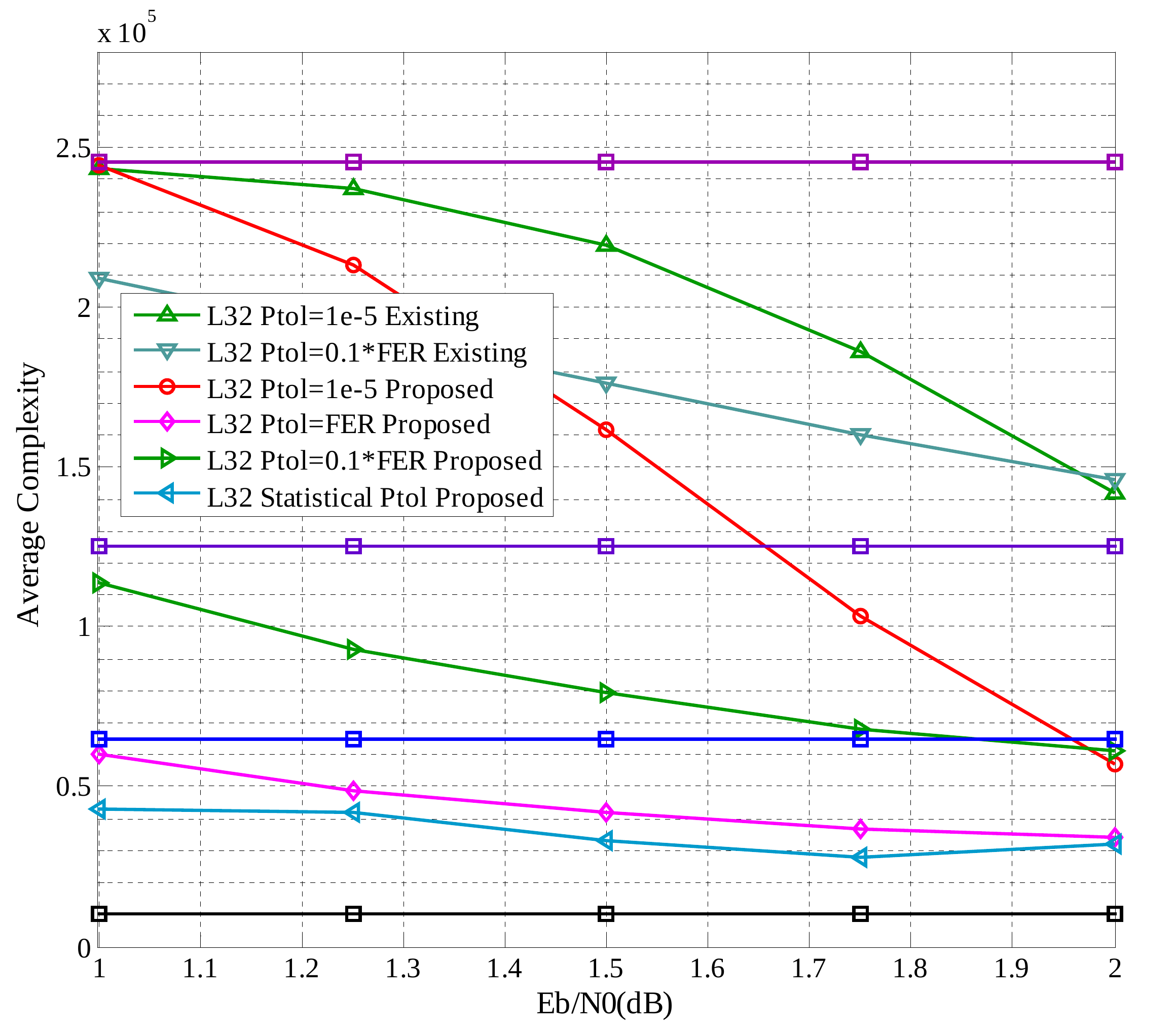}}
\caption{Average computational complexity.}
\label{fig_complexity}
\end{figure}

\begin{figure}
\centering{
\includegraphics[width=0.85\columnwidth]{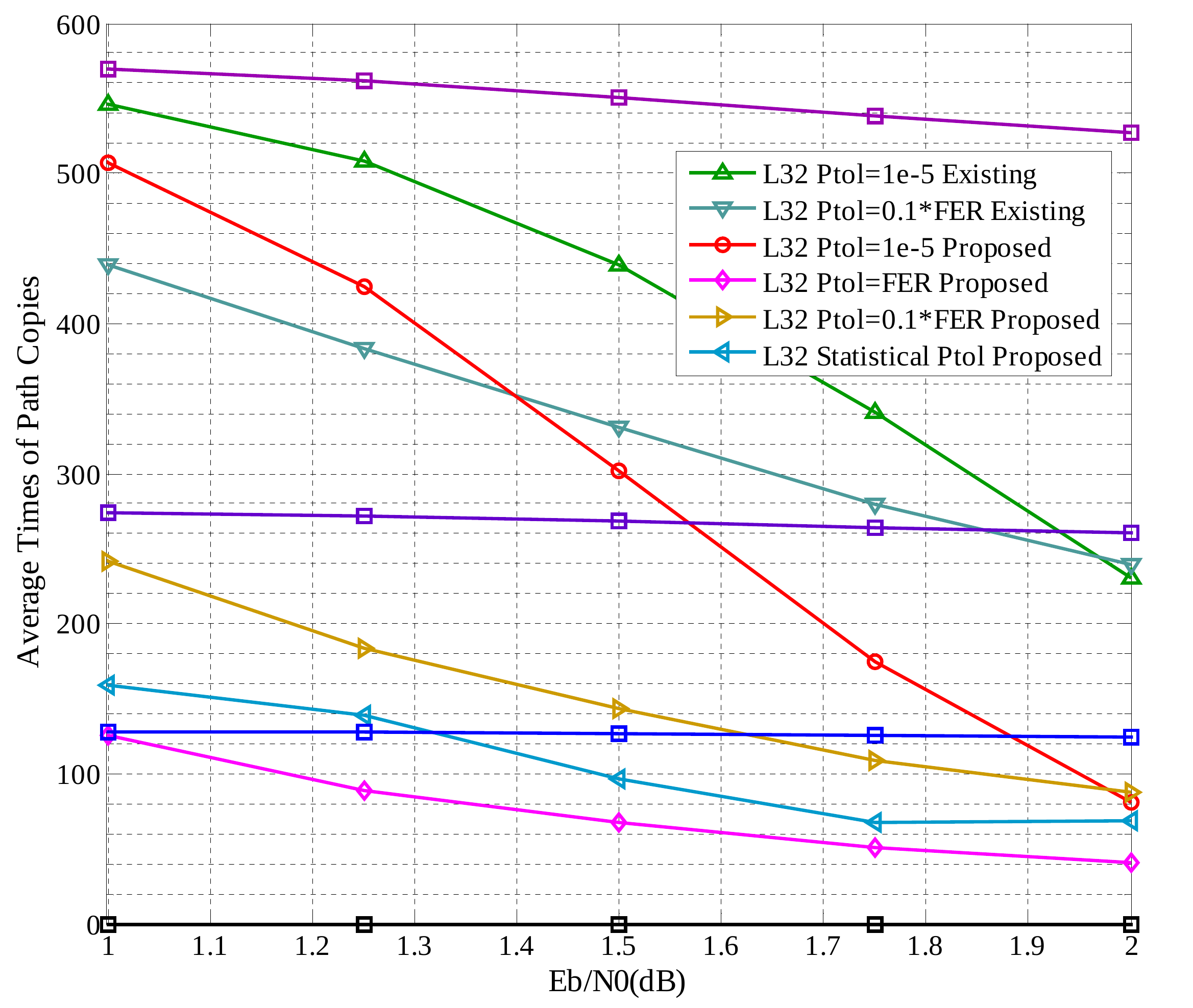}}
\caption{Number of path copy operations.}
\label{fig_pathcopy}
\end{figure}

Fig. \ref{fig_bler} shows the FER performances under different $L$ values and pruning techniques.
Fig. \ref{fig_complexity} and Fig. \ref{fig_pathcopy} show the corresponding average computational complexity and average number of path copies, respectively.
The average computational complexity is evaluated in terms of the number of metric recursive operations, which are defined in (\ref{equ_app_recursive1}) and (\ref{equ_app_recursive2}).
Here, we pay more attention to the SNR region where the FER is around $0.1 \sim 0.001$, which is the interesting
Particularly, the thresholds of `sum statistical' is obtained by Monte Carlo simulation.
As shown in the figures, when $P_\text{tol}$ takes a relative conservative value (compared with the FER), that is $10^{-5}$, all the pruning technique do not introduce noticeable loss in FER; while the proposed scheme has much lower complexity than the existing scheme in \cite{vtc}.
When decoding with \mbox{CA-SCL} with $L=32$ and $P_\text{tol}=\text{FER}$, the performance is deteriorated and very close to standard \mbox{CA-SCL} with $L=16$, while the average complexity is even lower than the standard one with $L=8$.
Further, when $P_\text{tol}=0.1 \times \text{FER}$, the FER performance loss is less than $0.01$dB, but the complexity is reduced by $50\% \sim 75\%$.

\begin{figure}
\centering{
\includegraphics[width=0.85\columnwidth]{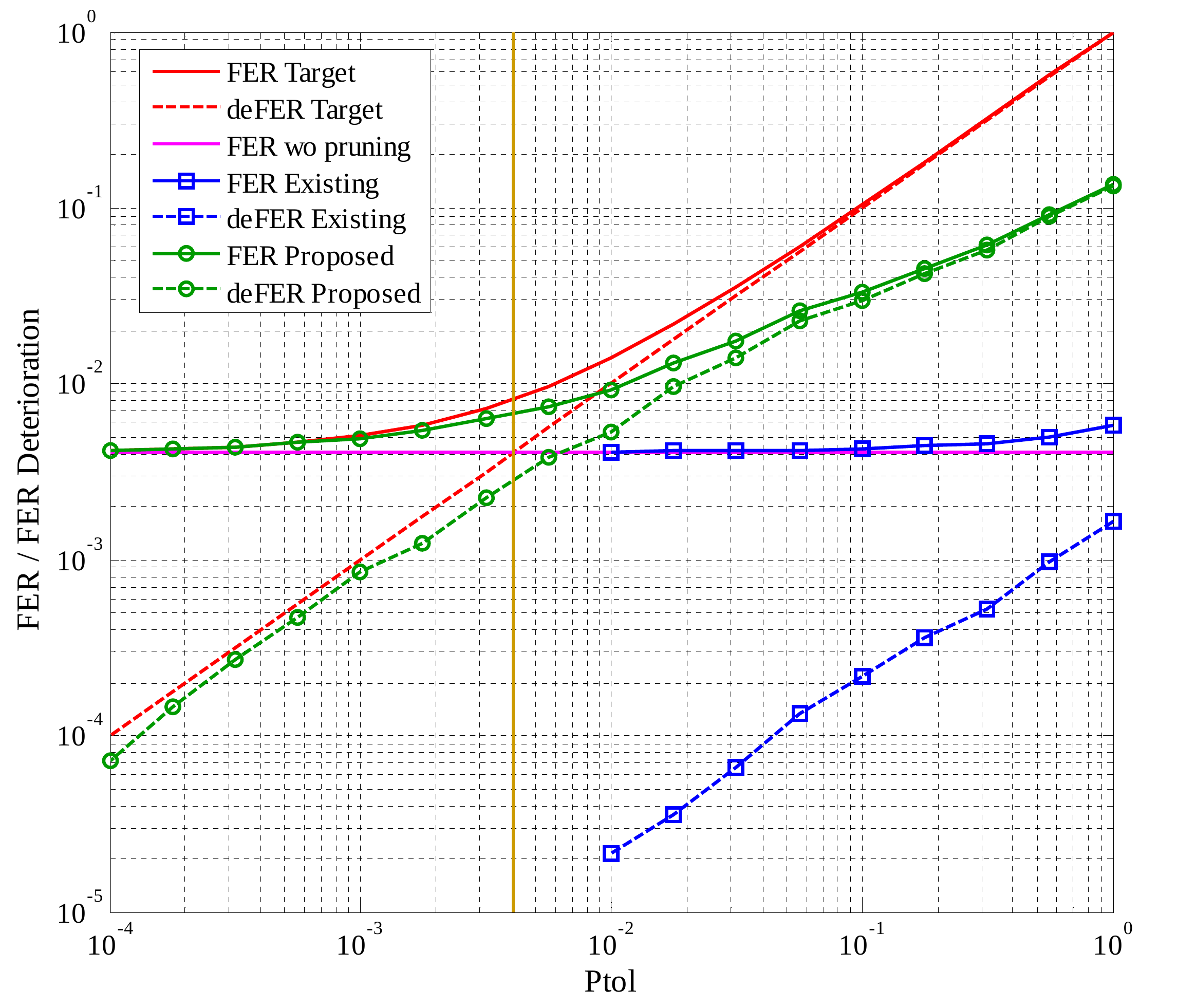}}
\caption{FER performance deterioration comparison of the pruned decoding scheme.}
\label{fig_debler}
\end{figure}

Fig. \ref{fig_debler} compares the FER and FER loss of the proposed pruning scheme and \cite{vtc} under different target losses $P_\text{tol}$.
The $E_b/N_0$ is fixed to $1.5$dB.
As shown in the figure, when $P_\text{tol} \le \text{FER}$, the actual FER loss is very close to the target $P_{tol}$; while the actual loss of \cite{vtc} is far less than the target.
That means, compared with \cite{vtc}, the proposed pruning scheme utilizes the tolerant FER loss much more efficiently, thus it is with lower complexity.

\section{Conclusion}
\label{sec_con}

In this paper, a tree-pruning technique to reduce the complexity of (CA-)SCL is proposed.
During the decoding process, the candidate paths with metric less than a threshold are directly deleted to avoid redundant path extensions.
Based on the reliabilities of the information/frozen bits, an upper bound of the path metric is derived to estimate the deterioration brought by the pruning operation.
Utilizing this bound, a dynamic thresholding technique is presented.
Compared with a similar existing scheme \cite{vtc}, the new proposed scheme can make full use of the given tolerant performance deterioration, and is much more efficient.

\end{document}